\documentclass[aps,prb,twocolumn,showpacs,psfig,superscriptaddress,longbibliography]{revtex4-1}
\usepackage{textcomp}
\usepackage{times}
\usepackage{graphicx}
\usepackage{float}
\usepackage{latexsym,amsmath,amssymb,bm,euscript}
\usepackage{color}
\usepackage{subfigure}
\usepackage{epstopdf}
\usepackage[colorlinks=true,linkcolor=blue,citecolor=blue]{hyperref}
\usepackage{hyperref}
\usepackage{soul}
\usepackage[normalem]{ulem}
\usepackage{mathrsfs}
\usepackage{amsmath}
\usepackage{lettrine}
\usepackage{xspace}
\usepackage{textcomp}
\usepackage{xr}
\usepackage{appendix}
\usepackage{verbatim}
\usepackage{longtable}

\begin{document}

\title{Haldane phase, field-induced magnetic ordering and Tomonaga-Luttinger liquid behavior
in a spin-one chain compound NiC$_2$O$_4$$\cdot$2NH$_3$}

\author{Shuo Li}
\thanks{These authors contributed equally to this study.}
\affiliation{School of Physics and Beijing Key Laboratory of Opto-electronic Functional Materials $\&$ Micro-nano Devices, Renmin
University of China, Beijing 100872, China}
\affiliation{Institute of Physics, Chinese Academy of Sciences, and Beijing National Laboratory for Condensed Matter Physics, Beijing 100190, China}

\author{Zhanlong Wu}
\thanks{These authors contributed equally to this study.}
\affiliation{School of Physics and Beijing Key Laboratory of
Opto-electronic Functional Materials $\&$ Micro-nano Devices, Renmin
University of China, Beijing 100872, China}

\author{Yanhong Wang}
\thanks{These authors contributed equally to this study.}
\affiliation{School of Chemistry and Chemical Engineering, Huazhong University of Science and Technology, Wuhan, 430074, China}

\author{Jun Luo}
\affiliation{Institute of Physics, Chinese Academy of Sciences, and Beijing National Laboratory for Condensed Matter Physics, Beijing 100190, China}

\author{Kefan Du}
\affiliation{School of Physics and Beijing Key Laboratory of
Opto-electronic Functional Materials $\&$ Micro-nano Devices, Renmin
University of China, Beijing 100872, China}

\author{Xiaoyu Xu}
\affiliation{School of Physics and Beijing Key Laboratory of
Opto-electronic Functional Materials $\&$ Micro-nano Devices, Renmin
University of China, Beijing 100872, China}

\author{Ze Hu}
\affiliation{School of Physics and Beijing Key Laboratory of
Opto-electronic Functional Materials $\&$ Micro-nano Devices, Renmin
University of China, Beijing 100872, China}

\author{Ying Chen}
\affiliation{School of Physics and Beijing Key Laboratory of
Opto-electronic Functional Materials $\&$ Micro-nano Devices, Renmin
University of China, Beijing 100872, China}

\author{Jie Yang}
\affiliation{Institute of Physics, Chinese Academy of Sciences, and Beijing National Laboratory for Condensed Matter Physics, Beijing 100190, China}

\author{Zhengxin Liu}
\affiliation{School of Physics and Beijing Key Laboratory of
Opto-electronic Functional Materials $\&$ Micro-nano Devices, Renmin
University of China, Beijing 100872, China}
\affiliation{Key Laboratory of Quantum State Construction and Manipulation (Ministry of Education),
Renmin University of China, Beijing 100872, China}

\author{Rong Yu}
\affiliation{School of Physics and Beijing Key Laboratory of
Opto-electronic Functional Materials $\&$ Micro-nano Devices, Renmin
University of China, Beijing 100872, China}
\affiliation{Key Laboratory of Quantum State Construction and Manipulation (Ministry of Education),
Renmin University of China, Beijing 100872, China}

\author{Yi Cui}
\email{cuiyi@ruc.edu.cn}
\affiliation{School of Physics and Beijing Key Laboratory of
Opto-electronic Functional Materials $\&$ Micro-nano Devices, Renmin
University of China, Beijing 100872, China}
\affiliation{Key Laboratory of Quantum State Construction and Manipulation (Ministry of Education),
Renmin University of China, Beijing 100872, China}

\author{Rui Zhou}
\email{rzhou@iphy.ac.cn}
\affiliation{Institute of Physics, Chinese Academy of Sciences, and Beijing National Laboratory for Condensed Matter Physics, Beijing 100190, China}

\author{Hongcheng Lu}
\email{hcl@hust.edu.cn}
\affiliation{School of Chemistry and Chemical Engineering, Huazhong University of Science and Technology, Wuhan, 430074, China}

\author{Weiqiang Yu}
\email{wqyu\_phy@ruc.edu.cn}
\affiliation{School of Physics and Beijing Key Laboratory of
Opto-electronic Functional Materials $\&$ Micro-nano Devices, Renmin
University of China, Beijing 100872, China}
\affiliation{Key Laboratory of Quantum State Construction and Manipulation (Ministry of Education),
Renmin University of China, Beijing 100872, China}


\begin{abstract}
We performed single-crystal magnetic susceptibility and $^1$H NMR measurements
on a  quasi-1D, spin-1 antiferromagnet NiC$_2$O$_4$$\cdot$2NH$_3$, with temperature down to
100~mK and with field up to 26~T.
With field applied along the chain direction (crystalline $b$ direction), a spin gap is determined at low fields.
Our susceptibility and spin-lattice relaxation measurements  reveal a Haldane phase at low field,
with an intrachain exchange coupling $J$~$\approx$~35~K and an easy-plane single-ion anisotropy of 17~K.
A field-induced antiferromagnetic (AFM) ordering emerges at fields of 2.1~T, which sets a three-dimensional (3D) quantum critical point (QCP).
The high-temperature spin-lattice relaxation rates $1/T_1$ resolves
an onset of Tomonaga-Luttinger liquid behavior at field above $3.5$~T, which characterizes a hidden 1D QCP.

\end{abstract}

\maketitle

\section{\label{Intro} Introduction}

After Haldane's conjecture in 1983~\cite{1983_PRL_Haldane}, Heisenberg antiferromagnetic (HAFM) chains
with integer spins have attracted a lot of interests.
In contrast to half-integer spin chains with gapless, continuum excitations, the integer spin HAFM chains
are equivalent to a valence-bond-solid(VBS) state without  long-range magnetic ordering
and depicted by a hidden string order, representing the breaking of non-local
hidden symmetries~\cite{1989_PRB_Rommelse,1992_CMP_Tasaki,1993_PRB_Seiji}.
A closely related 1D, spin-1 chain with AFM
interaction was precisely solved by Affleck, Kennedy, Lieb, and Tasaki (AKLT)~\cite{1987_PRL_Affleck},
and Haldane phase can be adiabatically connected to the AKLT state~\cite{1987_PRL_Affleck},
which is a one-dimensional(1D) magnetic symmetry-protected-topological(SPT) phase containing
a bulk triplet gap $\Delta$ and a gapless edge mode~\cite{1983_PRL_Haldane,1994-PRB-TKNg}.
With application of external field above a critical value, a gapless Tomonaga-Luttinger liquid (TLL) is induced ~\cite{,1980_Schulz_PRB,2003_book_Giamarchi,2006_Sato_JOSM}.

In bulk materials, single-ion anisotropy ($D$) and interchain
couplings ($J'$) are inevitable, which could suppress the AKLT state and lead to other phases~\cite{1986_PRB_Schulz,2013_PRB_Law}.
As presented by a phase diagram in the $D$-$J'$ plane, the Haldane phase is confined in a limited parameter space,
neighboring AFM and gapped large-$D$ phases~\cite{2021_Scireport_Kozlyakova}.
Up to now, several nicklate magnets were reported as the Haldane systems, notably
Ni(C$_2$H$_8$N$_2$)$_2$NO$_2$ClO$_4$ (NENP)~\cite{1986_PRL_Buyers,1987_EurophysicsLwtt_stirling,1989_JoAP_vettier,1989_PRL_Toshiya,1990_JPSJ_Goto,1992_PRB_Kohmoto,1992_PRL_Renard,1997_PhysB_Regnault,1999_Regnault_book},
Ni(C$_5$H$_{14}$N$_2$)$_2$N$_3$(PF$_6$) (NDMAP)~\cite{1997_JOP_Yamashita,1998_PRL_Katsumata}, Y$_2$BaNiO$_5$~\cite{1995_PRB_Cheong,1996_PRB_Aeppli}, and some others
~\cite{1990_JOMMM_Verdaguer, 1991_PRB_Veillet,1995_JOMMM_Renasrd,2001_PRB_harada,2005_PRB_Takano,2022_NC_ZLXue}.
In NENP, the spin gap~\cite{1986_PRL_Buyers,1987_EurophysicsLwtt_stirling,1990_JPSJ_Goto,1991_PRL_Lu,1992_PRB_Kohmoto,1992_PRL_Renard,1997_PhysB_Regnault}
and a field-induced staggered transverse magnetization~\cite{1991_PRB_Morimoto} were reported.
Experimental verifications of Haldane phase and its novel excitations are still demanding in the context of topological phases.

The Hamiltonian for the intrachain couplings of quasi-1D, AFM spin-1 materials is expressed as~\cite{2017_PRB_Tzeng},
\begin{equation}
H=\sum_{i}J\vec{S}_{i}{\cdot}\vec{S}_{i+1}+D(S_i^z)^2+E[(S_i^x)^2-(S_i^y)^2],
 \label{equa1}
\end{equation}
where $J$ ($>$0) represents exchange couplings, $D$ and $E$ represent single-ion anisotropy which
are caused by spin-orbit coupling and crystal field.
The Haldane phase exists in a range of $D$ values between $-0.29J\le D\le 0.99J$~\cite{1991_PRB_Minour,1992_PRB_Lacaze_2}.
For $D$~$>$~$0.99J$, the ground state is the large-$D$ state which is gapped and topologically trivial~\cite{2021_SSMS_Vasiliev}.

For the isotropic limit ($D$~$=$~$E$~$=$~0), the Haldane phase is characterized by triplet excitations
with a minimum gap $\Delta$~$=$~$0.41J$ at $q$~$=$~$\pi$ and a gap 2$\Delta$ at $q$~$=$~$0$, above which two-particle continuum emerges~\cite{1989_PRL_santos,1993_PRB_David}.
The addition of the $D$ term splits the triplet excitations into anisotropic excitations, including a singlet with  a
gap $\Delta_z$  and a doublet with a gap $\Delta_{xy}$~\cite{1992_PRB_Lacaze_1,1992_PRL_Renard}.
For $0\le D \le 0.25J$, these two gaps scale linearly with $D$ by~\cite{1992_PRB_Lacaze_1,1992_PRL_Renard}
\begin{equation}
\Delta_z = 0.41 J + 1.41 D, \\
\Delta_{xy} = 0.41 J- 0.57 D.
\label{equa2}
\end{equation}
$\Delta_{xy}$ will  further split with the in-plane anisotropy term $E$.

Recently, a quasi-1D, spin-1 compound NiC$_2$O$_4$$\cdot$2NH$_3$ (NiCO) was reported~\cite{2022_JSSC_HCLu} .
In this paper, we report single-crystal $^1$H NMR measurements down to 100~mK and with
fields up to 26~T, and susceptibility measurements down to 2~K.
With field applied along the crystalline $b$ axis (chain direction),
we found that the system follow the Haldane physics with an easy-plane single-ion anisotropy $D$ and
negligible in-plane anisotropy $E$.

The ground state is gapped with $\Delta_{xy}=4.9$~K, which closes at a 3D critical field $H^{\rm 3D}_{\rm c}$~$\approx$~2.1~T.
Beyond $H^{\rm 3D}_{\rm c}$, a glassy behavior is observed  and
a field-induced magnetic ordering occurs at low temperatures with magnetic field above $H^{\rm AFM}_{\rm c}$(2.44~T).
With field above 3.5~T, a power-law scaling between $1/T_1$ and temperature is found above $T_{\rm N}$,
which characterizes a TLL liquid behavior and a hidden 1D critical field $H^{\rm 1D}_{\rm c}$ (3.5~T).
The model parameters of the system are then determined as $J$~$\approx$~35~K, $D$~$\approx$~$0.47(4)J$, $E \approx 0$ and $J'$~$\approx$~$0.022J$.

The paper is organized as the following. Materials and experimental methods are presented in Sec.~\ref{struc}.
In Sec.~\ref{S_susceptibility}, magnetic susceptibility  of NiC$_2$O$_4$$\cdot$2NH$_3$ measured with different field orientations are analyzed.
In Sec.~\ref{S_spectra}, NMR spectra at typical fields are shown, which resolves the distinction between the gapped Haldane phase and
the ordered AFM phase. In Sec.~\ref{relaxation}, $1/T_1$ are reported,
where the Haldane gap $\Delta$, the N\'{e}el temperature $T_{\rm N}$, and  Luttinger exponent $\eta$ are determined.
Finally, an ($H$, $T$) phase diagram is established and shown in Sec.~\ref{S_phase}.

\section{\label{struc} Materials and Methods}

\begin{figure}[t]
\includegraphics[width=8.5cm]{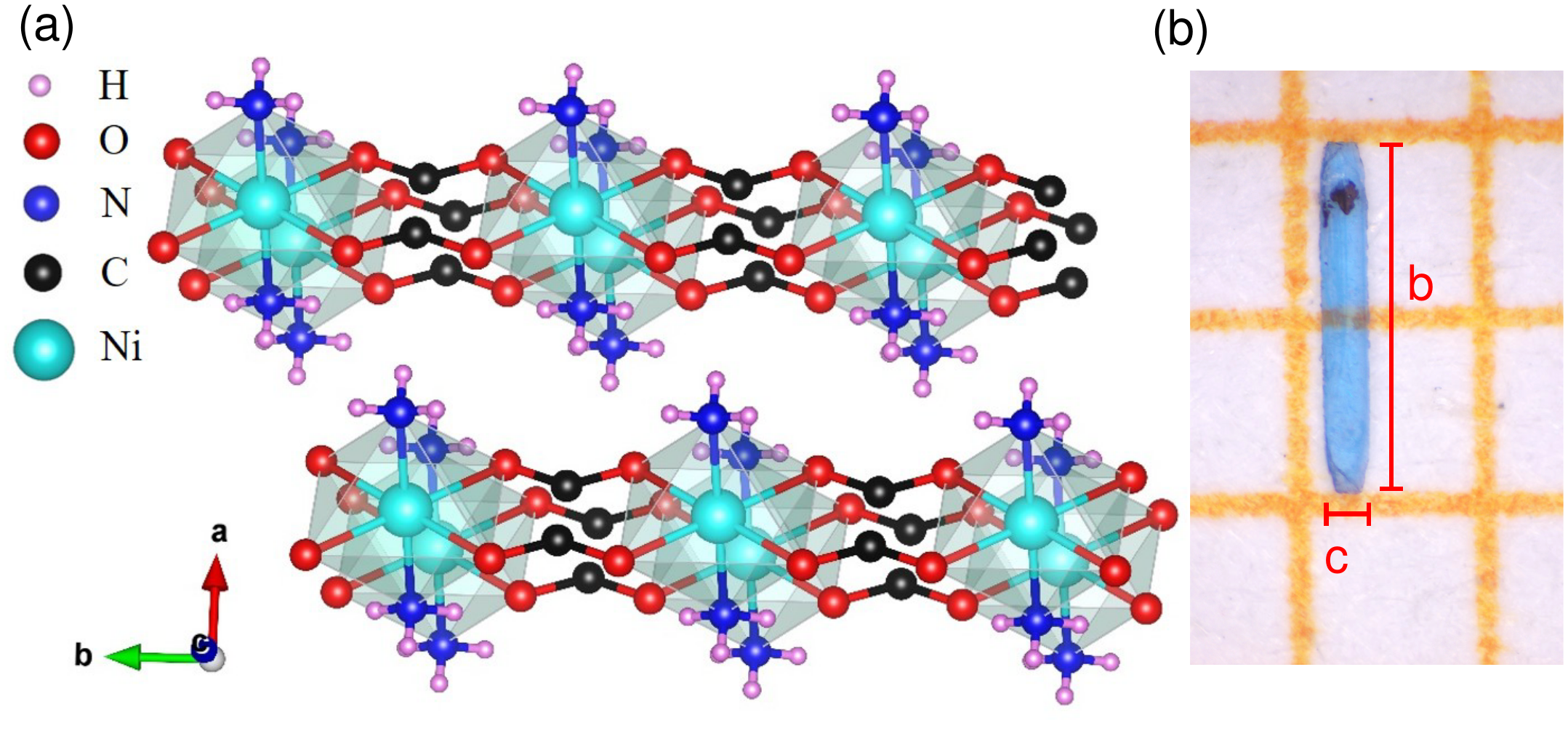}
\caption{\label{struc1}
\textbf{Lattice structure of NiCO.}
(a) Lattice structure viewed along the $c$ axis. NiO$_4$N$_2$ octahedra and shared C$_2$O$^{2-}_4$ unit are aligned in parallel
along the $b$ axis (the chain direction). NH$_3$ groups are located above and below the Ni$^{2+}$ ions.
(b) Picture of a NiCO single crystal with the $b$ axis and $c$ axis marked.
}
\end{figure}

NiCO crystallizes in a centrosymmetric monoclinic structure in the space group \textit{C2/m} as shown in Fig.~\ref{struc1},
with lattice parameters $a=10.767$~\AA, $b=5.414$~\AA ~and $c=5.005$~\AA~\cite{2022_JSSC_HCLu}.  
The material contains NiC$_2$O$_4$ octahedra, with Ni$^{2+}$  strongly coupled along the crystalline $b$ axis
to form the spin-1 chain.

The NiCO single crystals were first synthesized using the hydrothermal method by Kenneth R. Poeppelmeier, who discovered that the material exhibits magnetic properties ~\cite{2022_JSSC_HCLu}.
High-quality single crystals with typical dimensions of 0.5$\times$0.2$\times$0.1 mm$^3$
are chosen for the measurements. Bulk susceptibility were measured in a Quantum-Design magnetic property measurement
system (MPMS) with field along different crystalline directions.
NMR measurements were performed on $^1$H nuclei (gyromagnetic ratio $\gamma$$=$42.5759~MHz/T).
The sample was fixed in a silver coil with Cytop glue which is proton free,
and cooled in a Variable Temperatures Insert (VTI) with temperature down to 1.8~K and in a
dilution refrigerator with temperature down to 100~mK.
NMR measurements above 16~T were performed Synergetic Extreme Condition User Facility (SECUF),
with the 26T all-superconducting magnet.

$^1$H NMR spectra were collected by the spin-echo method, using $\pi/2$-$\tau$-$\pi/3$ sequences, where $\pi/2$ and $\pi/3$ denote
typical RF pulses for amino group, affected by proton rotation, with time durations of 5~$\mu$s and 3~$\mu$s, respectively.
For broad spectra at high fields, the spectra were obtained by sweeping frequencies at fixed field.
The NMR Knight shift was calculated by $K_{\rm n}$$=$$(f/\gamma H$$-$$1)$$\times$$100\%$, where $f$ is the
average frequency of the whole spectrum.

The spin-lattice relaxation rates 1$/T_1$ was measured by the spin inversion-recovery method.
$T_1$ was obtained by fitting the nuclear magnetization $m(t)$ to the recovery function $m(t)=m(\infty)[1-ae^{-(t/T_1)^{\beta}}]$,
where $\beta$ is a stretching factor. $\beta$ is found to be about 1 in the paramagnetic phase, which indicates high-quality of the sample.

\section{\label{S_susceptibility} Bulk susceptibility}

\begin{figure}[t]
\includegraphics[width=8.5cm]{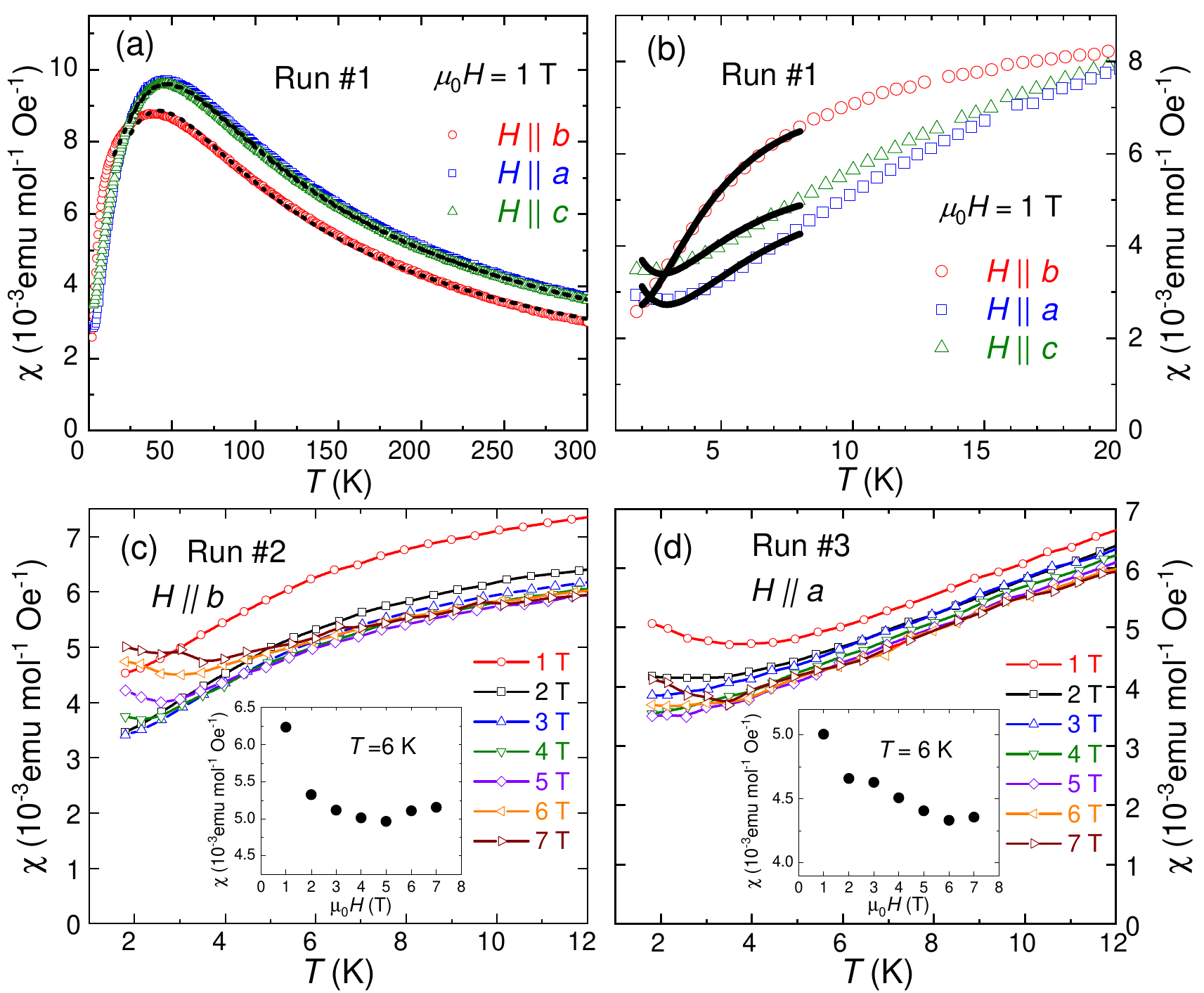}
\caption{\label{susceptibility2}
\textbf{Magnetic susceptibility.}
(a) $\chi$  as functions of temperatures at a field of 1~T, with the field applied along the crystalline $a$, $b$, and $c$ axis, respectively.
The dash lines represent high-temperature fits to Eq.~\ref{equa3}, covering the temperatures from 300~K down to 30~K.
(b) Enlarged view of the low-temperature data.
 The solid lines represent fits to Eq.~\ref{equa5}.
In (c) \& (d),  $\chi(T)$ is depicted  as functions of temperatures for different fields applied along the $a$ and $b$ axis, respectively. Insets: $\chi$ as functions of fields at a constant temperature 1.8~K.
Data for each panel were taken during separate experimental runs, as indicated in each panel.
}
\end{figure}

The low-field magnetic susceptibility $\chi(T)$ were measured under a field of 1~T, applied along three crystalline axes respectively,
with data shown in Fig.~\ref{susceptibility2}(a).
Upon cooling, $\chi(T)$ first increases and exhibits a broad peak around 45~K with field along the $a$ and the $c$ axis,
and at 37~K with field along the $b$ axis, which is a typical behavior of 1D magnets~\cite{1964_PR_Bonner}.
At temperature above 50~K, $\chi$ shows nearly identical values with field along the $a$ and $c$ axis, but larger
than that along the $b$ axis, which suggests the presence of an easy-plane anisotropy $D$, whereas the
in-plane anisotropy term $E$ is likely very small.

We first attempt to fit the high-temperature $\chi$ to the form for an isotropic Haldane chain~\cite{1982_Chemistry_olivier,2022_JSSC_HCLu},
  \begin{equation}
\begin{aligned}
\chi=\frac{N\mu_{\rm B}^2g^2}{k_{\rm B}T}\frac{2+0.0194x+0.777x^2}{3+4.346x+3.232x^2+5.834x^3},
\label{equa3}
\end{aligned}
\end{equation}
where $x$$=$$J$/$k_{\rm B}T$, and $N$, $\mu_{\rm{B}}$, $g$, and $k_B$ are Avogadro constant, Bohr magneton, $g$-factor
and Boltzmann constant, respectively. All fits, as demonstrated in Fig.~\ref{susceptibility2}(a),
are succeeded with temperatures from 30~K to 300~K.
The fitting parameters, with three field orientations, are obtained as
\begin{equation}
\begin{aligned}
\begin{split}
&J_a/k_B=35.6\pm0.1~{\rm K}, g_a=2.30\pm0.01,\\
&J_b/k_B=31.8\pm0.1~{\rm K}, g_b=2.07\pm0.01,\\
&J_c/k_B=35.0\pm0.1~{\rm K}, g_c=2.27\pm0.01.
\end{split}
\label{equa4}
\end{aligned}
\end{equation}

The difference of $J_a$, $J_b$ and $J_c$ is within 15\%, therefore a large $D$ limit is ruled out~\cite{1992_PRB_Lacaze_2}.
Considering that the susceptibility along the $b$ axis is smaller than that along the $a$ axis, it is suggested that $D>0$.
With this, we use $J$~$=$~$(J_a$$+$$J_c)/2$~$\approx$~35.3~K as the dominant term of the system.


Below 20~K, the sharp drop of $\chi$ upon cooling is a signature of gap opening.
In fact, the data can be fit to combined contributions from both 1D gapped excitations~\cite{1999_Regnault_book}
and a Schottky-like term by
\begin{equation}
\chi\sim a T^{-0.5}exp(-\Delta/T)+n\tanh(\frac{\mu_{\rm{B}}B}{k_{\rm{B}}T}),
\label{equa5}
\end{equation}
where $\chi$, $\Delta$,$B$ and $n$ is magnetic susceptibility, energy gap, magnetic field and density of impurities respectively.
The Schottky-like term catches the upturn observed at temperature below 2~K, with $H||a$ and $H||c$, as illustrated in Fig.~\ref{susceptibility2}(b).
This upturn is likely due to the presence of magnetic impurities within the material.
The gaps are obtained as $\Delta_a=9.92$~K, $\Delta_b=7.10$~K, and $\Delta_c=8.41$~K at 1~T with three field orientations,
by fits in a temperature range from 2~K to 8~K, as shown in Fig.~\ref{susceptibility2}(b).
As $\chi$ is dominated by excitations at $q$~$=$~0, the Haldane gap is estimated to be $\Delta_{xy}$~$=$~$\Delta_b/2$~$\approx$~3.55~K at 1~T, whereas
the larger gap $\Delta_{z}$ may not be observable in $\chi$ at such low temperatures.
 Given that $\Delta_{xy}$~$=$~$0.41J$$-$$0.57D$ (see Eq.~\ref{equa1}), $D$ is estimated as 0.47~$J$, which indicates that the system belongs to
the Haldane class with an easy-plane anisotropy~\cite{1992_PRB_Affleck,1999_Regnault_book,2014_PRL_Pinaki,2014_MPLB_Pinaki}.
The value of $\Delta_{xy}$ and $D$ will be further confirmed by subsequent NMR measurements.

To elucidate the intrinsic behavior of the materials, $\chi$ are measured as functions of temperature,
under various magnetic field strengths and orientations, as illustrated in Fig.~\ref{susceptibility2}(c)-(d).
As the magnetic field increases, the upturn of magnetic susceptibility is diminished, resembling the behavior of free magnetic impurities.
We then plotted $\chi$ as a function of magnetic field, at temperature about 6~K where the upturn is negligibly small,
as shown in the inset of Fig.~\ref{susceptibility2}(c)-(d).
A non-monotonic behavior is then seen with fields both along $b$ and $a$ axis respectively,
which suggests a field induced critical behavior. With field along the $b$ axis, a
1D quantum critical point is suggested at approximately 4~T, where $\chi(H)$ exhibits a dip feature.
In fact, our spin-lattice relaxation rate $1/T_1$, as shown later, resolves a field-induced TLL behavior at fields above 3.5~T.

\section{\label{S_spectra}NMR Spectra}

In the following, the $^1$H NMR spectra are presented, as shown in Fig.~\ref{spec3}, which reveal a Haldane phase at low fields and
a field-induced AFM ordering at high fields, as described below.

\subsection{\label{spechaldane} Paramagnetic and Haldane phase}

\begin{figure*}[t]
\includegraphics[width=18cm]{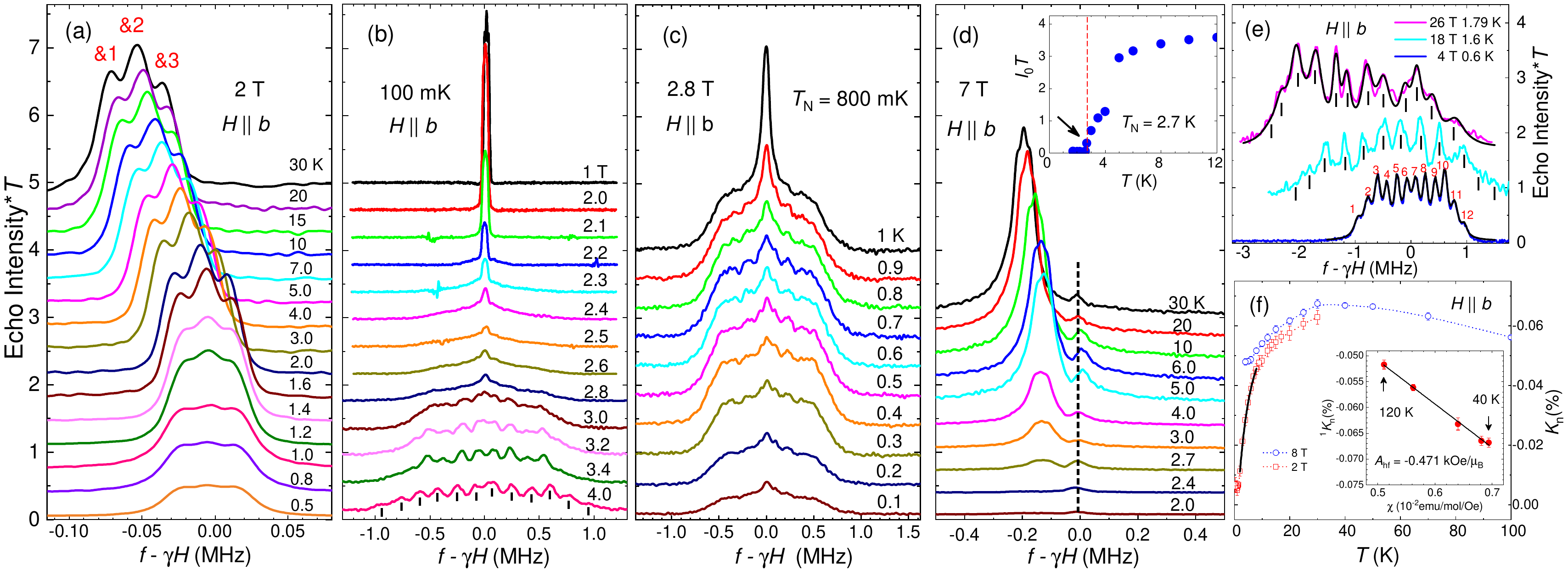}
\caption{{\label{spec3}}
\textbf{$^1$H NMR spectra with fields applied along the $b$ axis.}
(a) Spectra measured at a field of 2~T, with three peaks labeled as \&1, \&2, and \&3.
(b) Ultra-low temperature spectra with fields ranging from 1~T to 4~T. The positions of NMR peaks at 4~T are indicated by vertical lines.
(c) Spectra at 2.8~T, with $T_N$ determined to be 800~mK from the disappearance of the sharp central peak.
(d) Spectra measured at 7~T. The black dotted line points to a spurious $^1$H peak.
Inset: The integrated spectral weight as a function of temperature. The arrow marks $T_{\rm N}$,
which is set by a 90\% intensity reduction in this study.
(e) Spectra with fields of 4~T, 18~T and 26~T.
 Twelve NMR peaks are labeled by numbers (at 4~T) or the vertical lines (at 18~T and 26~T).
 The black solid lines, overlapping the 4~T and 26~T data, represent Lorentzian fits to the spectra, which include the twelve peaks.
(f) $K_ n$ plotted as a function of temperature at typical fields of 2~T and 8~T, below and above the field-induced phase transition.
Inset: $K_n$ as a function of $\chi$, measured at different temperatures and a constant field of 8~T, with
the hyperfine coupling constant obtained by a linear fit to the data.
}
\end{figure*}

The $^1$H NMR spectra under a field of 2~T applied along the $b$ axis, are shown with temperatures from 30~K down to 0.5 K, in Fig.~\ref{spec3}(a).
Three resonance peaks, marked by \&1, \&2 and \&3, can be distinguished at high temperatures, which correspond to three hydrogen positions within the NH$_3$ group.
The spectral width, as determined by the full-width at half-maximum (FWHM), measures approximately 51~kHz across the entire temperature range, although a smearing of three peaks is observed below 2~K.
The entire spectrum shifts toward higher frequencies with decreasing temperature below approximately 30~K.

\subsection{\label{speckn} Knight shift}

In fact, the Knight shift $K_n$, as depicted in Fig.~\ref{spec3}(f), is calculated from the
resonance frequency of the middle peak in the spectrum, which increases sharply as the temperature decreases.

As shown by the inset of Fig.~\ref{spec3}(f), $K_{\rm n}$ are
plotted against $\chi$, with data measured at various temperatures and a magnetic field of 8~T.
The data ranging from 40~K to 120~K conform to a linear fit, represented by the straight line.
From which, the hyperfine coupling is determined to be $A_{\rm{hf}}$~$\approx$~$-$0.471 kOe/$\mu_B$.

With such a negative hyperfine coupling,
the onset of a gapped phase is characterized by a sharp increase in the Knight shift.
With decreasing temperature,  $K_{n}$ first decreases and then increases when cooled below 30~K,
which is consistent with the change in the magnetic susceptibility, as depicted in Fig.~\ref{susceptibility2}(a).
In particular, $K_{n}$ at 2~T shows a very dramatic increase at temperatures below 20~K.
At a field of 8~T, $K_{n}$ approaches a finite value at the zero temperature limit, indicating the closure of the energy gap. This behavior is consistent with the presence of canted magnetization in the field-induced AFM phase.


\subsection{\label{specafm} High-field ordered AFM phase}

To precisely determine  the critical field that separates the Haldane phase from the ordered AFM phase,
the spectra were measured at 100~mK with increasing fields, as illustrated in Fig.~\ref{spec3}(b).
The spectra are narrow at low fields from 1~T to 2.1~T,
and a reduction in spectral intensity is observed at 2.1~T, which we identify as the critical magnetic field.
A broad shouldered feature appears at a field of 2.2~T,
with this shouldered feature serving as a signature of short-range AFM ordering.
At 2.3~T, the spectral weight at the shoulders grows larger than the center peak.
At 3~T, a broad spectrum is observed with several peaked features, indicating the presence of
different hyperfine fields on different hydrogen sites due to magnetic ordering.

The magnetic transition at high fields with varying temperature  can also be resolved.
As shown in Fig.~\ref{spec3}(c), spectra at a field of 2.8~T are measured over a temperature range
from 1~K down to 0.1~K. Upon cooling, the NMR spectrum is significantly broadened.
At 1~K, a central peak with a FWHM of approximately 0.1~MHz, which contains overlapped $^1$H contributions, is observed at the central position.
 However, broad shoulders are observed on both sides, indicating
the development of short-range AFM ordering.
 When cooled below 0.8~K, the sharp peak become barely resolvable,
and the FWHM of the whole spectrum reaches approximately 1~MHz, clearly indicating the onset of field-induced magnetic ordering.
Here we define $T_{\rm N}$($\approx$~0.8~K), which marks the emergence of long-range ordering and is characterized by the strong suppression of the central paramagnetic peak.
At a higher field of 7~T, the spectra measured at temperatures from 30~K to 2~K are displayed in Fig.~\ref{spec3}(d).
A single peak is observed down to 2~K, however, the spectral weight is clearly reduced
upon cooling, indicating the emergence of magnetic ordering.
For consistency, we define the N\'{e}el
temperature $T_{\rm N}$ as the temperature at which 90\% of the paramagnetic signal is lost, as illustrated in the inset of Fig.~\ref{spec3}(d).
The $T_{\rm N}$, determined by this scheme, is consistent with the value determined from the peaked feature in the $1/T_1$, as demonstrated in the subsequent section.

We further performed NMR measurements in the high-field magnet at SECUF, with fields of 18~T
and 26~T, as depicted in Fig.~\ref{spec3}(e).
Comparing the spectra in the ordered phase,
the 12-peak feature is consistently observed at all three fields, as indicated by the short vertical lines.

\begin{figure}[t]
\includegraphics[width=8.5cm]{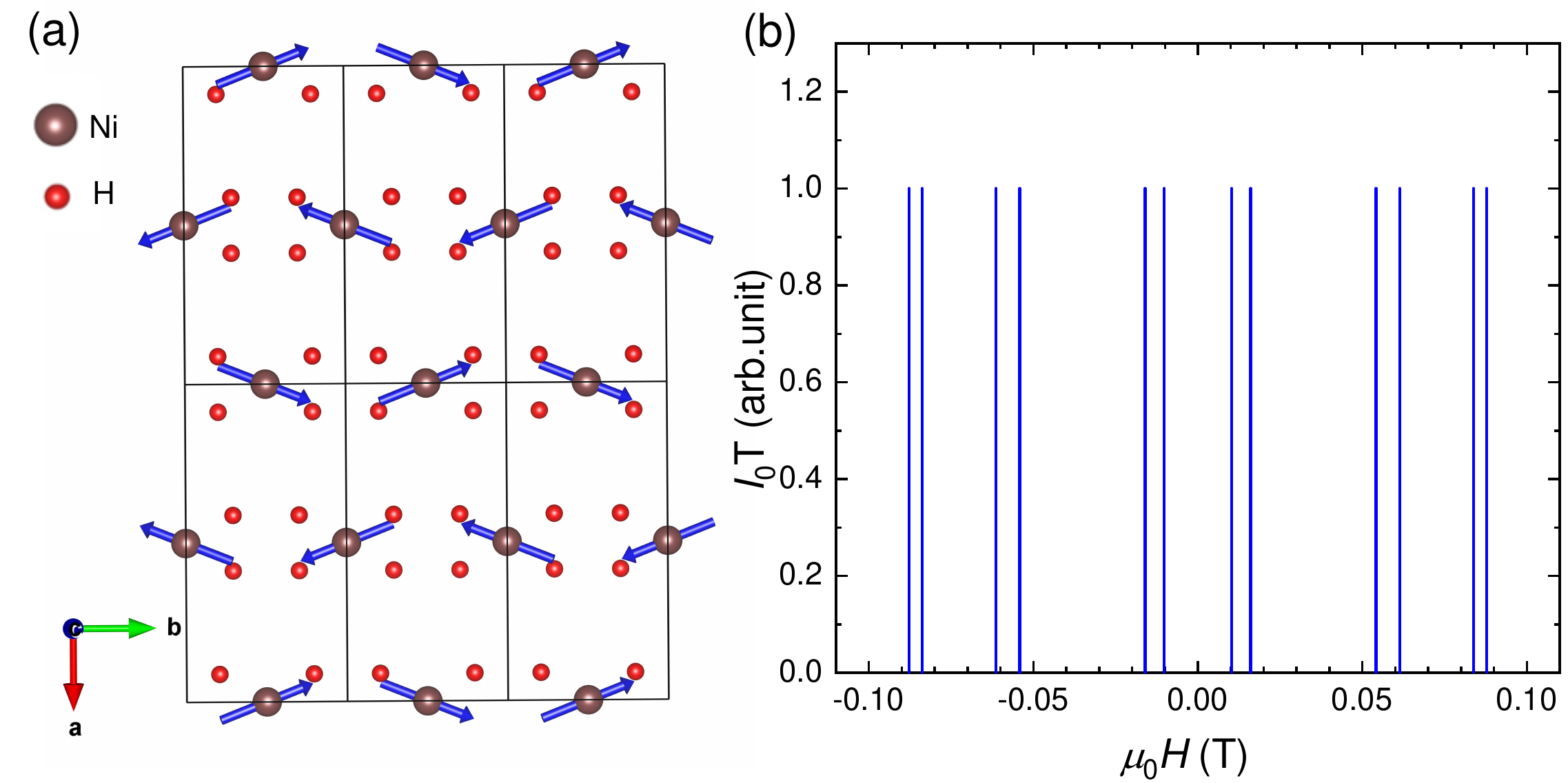}
\caption{{\label{spec4}}
\textbf{Spectral simulation of the field-induced AFM phase. }
(a) Proposed magnetic structure of the field-induced AFM phase.
(b) Simulated NMR spectra, assuming the magnetic structure shown in (a).
In the simulation, the moment size along the respective $a$, $b$, and $c$ directions is assumed with a ratio of 1:7:1
as an example.
}
\end{figure}

Indeed, the observed line splitting in the high-field, ordered phase is consistent with field-induced AFM ordering.  
We assume the magnetic structure depicted in Fig.~\ref{spec4}(a), where the AFM ordered moment are arranged in the crystalline $ac$ plane, perpendicular to the applied field ($b$ axis), and a canted moment along the field direction.
Using the dipolar-type hyperfine coupling, the NMR spectrum of this magnetic structure was simulated and presented in Fig.~\ref{spec4}(b).
The spectrum, which contains twelve peaks,  is consistent with the observations  shown in Fig.~\ref{spec3}(e).

\section{\label{relaxation}Spin-lattice relaxation rate}

The NMR spin-lattice relaxation rate $1/T_1$, with $1/T_1=\gamma^2 k_B T/ \mu_B^2 \sum_{q}A^2_{\rm{hf}}(q)$Im $\chi'(q,\omega)/\omega$,
probes  the low-energy fluctuation in the system,
where $\chi'(q,\omega)$ is dynamic susceptibility, $A_{\rm{hf}}(q)$ is hyperfine coupling constant, and $\omega$ is Larmor frequency.
In the following section, we present $1/T_1$  measured on $^1$H at various magnetic fields.

\subsection{\label{energygap} Low-field Haldane phase}

\begin{figure}[t]
\includegraphics[width=8.5cm]{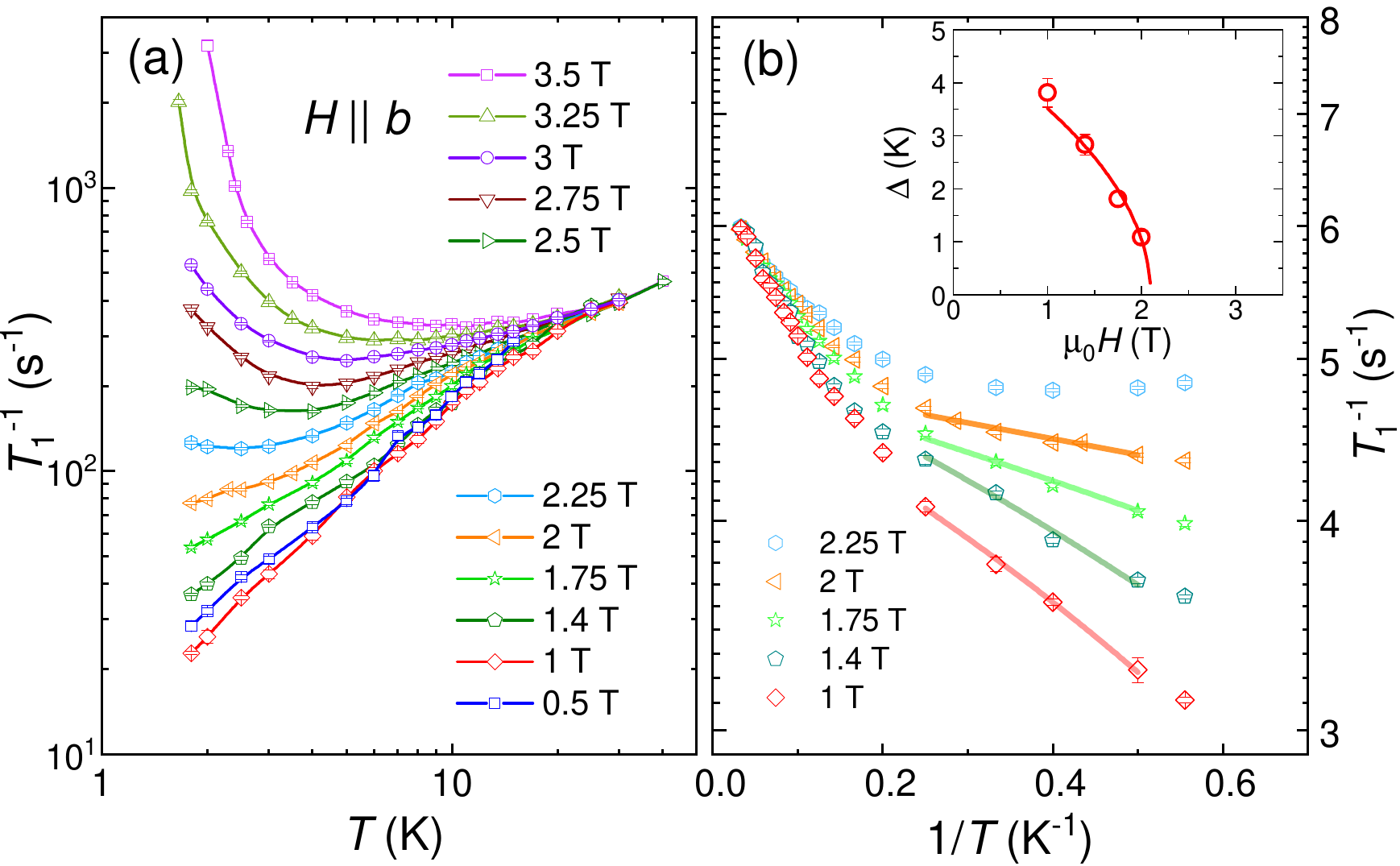}
\caption{{\label{invt14}}
\textbf{$\bm{1/T_1}$ at low fields. }
(a) 1$/T_1$ as functions of temperatures measured with field from 0.5~T to 3.5~T.
(b) Semilog plot of $1/T_1$ as functions of $1/T$ with field up to 2.25~T. The straight lines represent
gap function fits to the data at low temperatures (see text).
Inset: The obtained gap value $\Delta$ as a function of field.
The solid line is a function fit to $\Delta(H)$$\sim (H_{\rm c}-H)^{1/2}$ with field $H_{\rm c}=$~2.1~T.
}
\end{figure}

The $1/T_1$ measured at low fields along the $b$ axis, is presented  in Fig.~\ref{invt14} over a temperature range from 50~K to 1.8~K.
Upon cooling below 30~K, $1/T_1$ initially decreases rapidly, consistent with the 1D behavior observed at temperatures below $J$~\cite{1964_PR_Bonner}.
 In the low-field range from 0.5~T to 2~T, a monotonic decrease of $1/T_1$ with temperature indicates that the system remains in the gapped Haldane phase.
  Above fields of 2.25~T, an upturn is evident upon further cooling, suggesting that the Haldane phase is suppressed.
     The system then enters the field-induced TLL phase (above $T_{\rm N}$) and the ordered AFM phase (below $T_{\rm N}$) respectively.
     This phase will be discussed  in more detail later.

The spin gap in the Haldane phase can be deduced from the low-temperature behavior of $1/T_1$, which follows a gap function given by $1/T_1~{\propto}~e^{-\Delta/T}$.
The fitting is illustrated by a semilog plot of $1/T_1$ as a function of $1/T$,
where the low temperature data follows a straight line for each field below 2~T.
Indeed, such a fitting is only observed at temperatures significantly lower than $\Delta$, as suggested
by recent theoretical work~\cite{2019_PRB_Sengupta}.
The gap $\Delta$ at each field is then determined from
the slope of the line, with $\Delta$ depicted as a function of field in the inset of Fig.~\ref{invt14}(b).
Indeed, $\Delta(H)$ follows a critical behavior described by $\Delta(H)\sim(H_{\rm c}-H)^{1/2}$ with $H_{\rm c}~=~$2.1~T, as seen in the function fit of the figure.
Such a strong suppression of the gap with increasing field deviates from a linear behavior, which suggests that the system is influenced  by the interchain coupling, particularly when it is close to the field-induced magnetic ordering.

\subsection{Field-induced AFM ordering}

\begin{figure}[t]
\includegraphics[width=8.5cm]{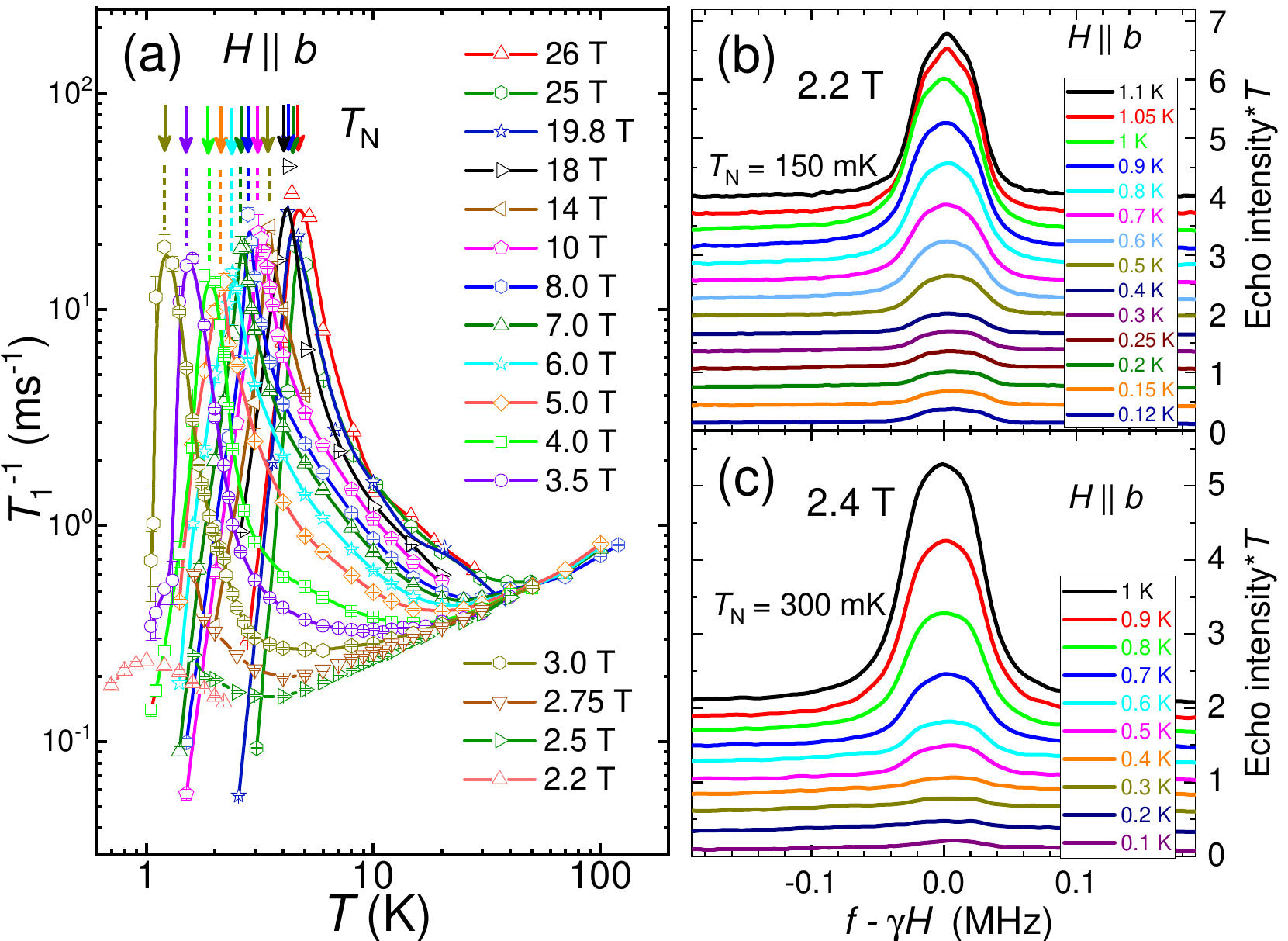}
\caption{{\label{invt16}}
\textbf{$\bm{1/T_1}$ at high fields. }
(a) $1/T_1$ as functions of temperatures. The arrows indicate the peak feature in 1/$T_1$, which is used to determine
the $T_{\rm N}$ as labeled.
(b)-(c) Spectra at different temperatures, under field of 2.2~T and 2.4~T, respectively.
The respective $T_{\rm N}$ at each field is defined as a temperature with 90\% loss of paramagnetic spectral weight.
}
\end{figure}

The $1/T_1$ with fields ranging from 2.2~T to 26~T are presented in Fig.~\ref{invt16}(a) as functions of temperature.
 The measurement were also taken on the middle peak of the spectra for each field.
For all these field, a peaked behavior at low temperatures is observed in $1/T_1$, which indicates the onset of field-induced magnetic ordering.
The peak position corresponds to $T_{\rm N}$, which is  produced
by strong low-energy spin fluctuations.

At fields of 3~T and above, a pronounced peak is observed in $1/T_1$, which clearly characterizes the
the phase transition temperature $T_N$, followed by the AFM ordering upon further cooling.
The sharp drop in $1/T_1$ below $T_{\rm N}$ indicates spin-wave excitations.
With increasing field up to 26~T, 
$T_{\rm N}$ shifts to higher temperatures.

With field ranging from 2.2~T to 3~T, the peak feature in $1/T_1$ is not prominent.
Then we performed a spectral weight analysis to determined $T_{\rm N}$, as illustrated in Fig.~\ref{invt16}(b) and (c).
At a field of 2.2~T, when the sample is cooled from 1.0~K to 0.1~K, a loss of spectral weight is observed at the center peaked, which suggests the onset of very slow spin fluctuations or a glassy behavior that suppresses the spectra before reaching the ordered phase.
For consistency, we define $T_{\rm N}$ as the temperature at which a 90\% loss of spectral weight is observed at the center peak, which corresponds to 0.15~K at this field.
A similar loss of spectral weight is observed at other fields above the critical field of 2.1~T,
and the corresponding $T_{\rm N}$ at 2.4~T is determined to be 0.3~K as shown in Fig.~\ref{invt16}(c).

Subsequently, all the $T_{\rm N}$ are obtained by these two methods
and are depicted in the phase diagram shown in Fig.~\ref{phase}(a).
The 3D quantum critical point between Haldane phase and AFM ordering
is determined to be $H^{\rm 3D}_{\rm c}$~$\approx$~2.1~T, which is lower than the 1D QCP (3.5~T).
 This 1D QCP  is determined by the high-temperature behavior of $1/T_1$ and will be further analyzed later.

For quasi-1D $S$~=~$1$ HAFM chain, $T_N$ can be approximated by the formula
$T_{\rm N}$~$\approx$~$(8J|J'|)^{1/2}$~\cite{1999_Regnault_book}.
Applying this to NiCO,
with $T_{\rm N} = 4$~K at a field far above the critical field,
the interchain coupling $J'$ is estimated to be approximately 0.77~K.
$T_{\rm N}$ increase monotonically with the field up to 26~T, indicating that the fully polarized phase
is at a very high field  that was not achieved in the current study.
 Using the above parameters, the QCP for the fully polarized phase is estimated
to be $H_{\rm c2}$~$\approx$~99.5~T, 
 deduced by $H_{\rm c2}$~$\approx$~$4J/g \mu_B$~\cite{1985_PRB_Parkinson,1991_PRB_Sakai}.

\subsection{High-field Tomonaga-Luttinger liquid behavior}

\begin{figure}[t]
\includegraphics[width=8.5cm]{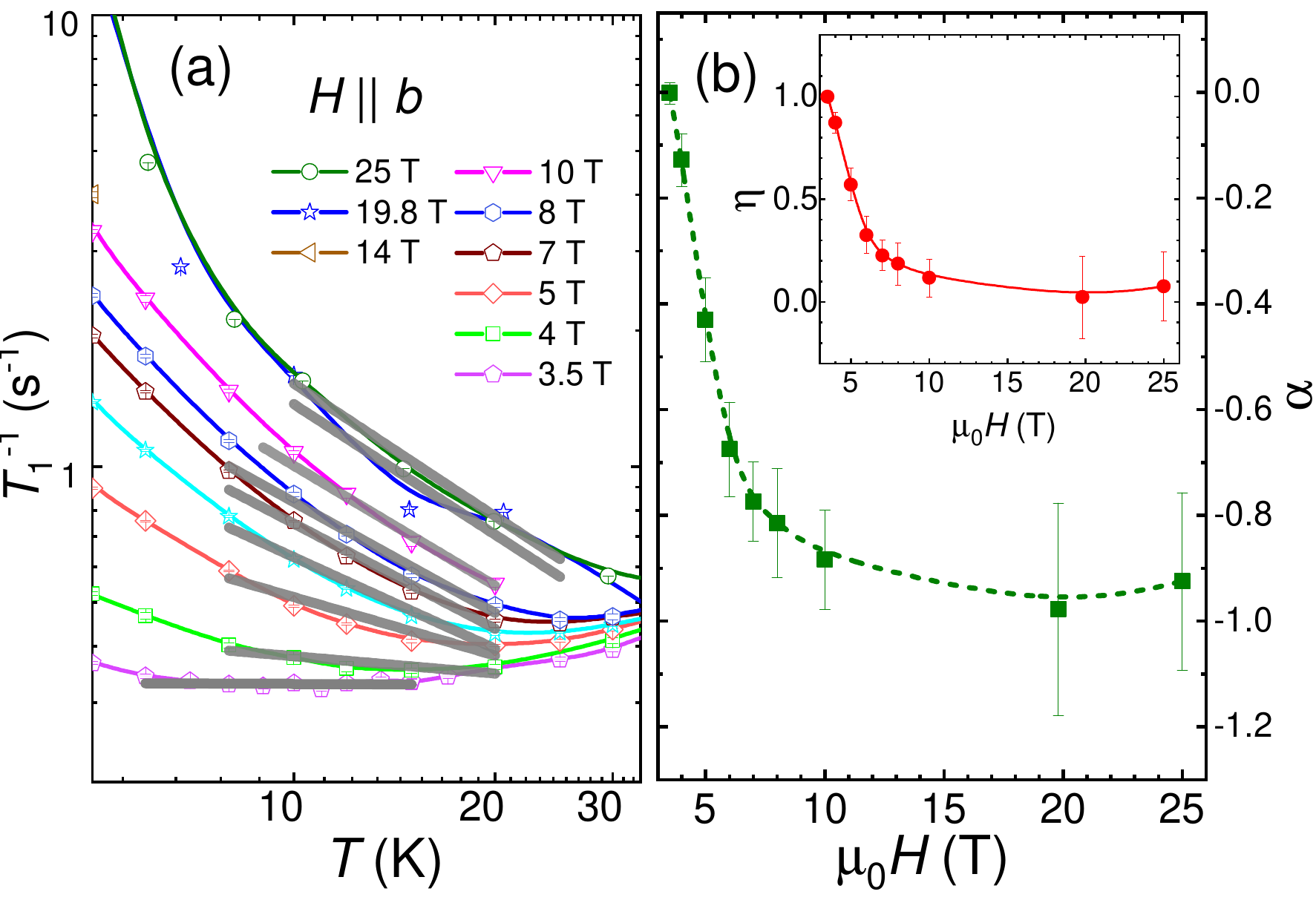}
\caption{{\label{invt17}}
\textbf{Spin-lattice relaxation rates in high-field PM phase. }
(a) An enlarge view of 1/$T_1$ data between 4~K and 30~K.
The straight gray lines represent power-law fits (see text) to the data far above $T_{\rm N}$ (see text).
(b) Power law exponent $\alpha$ at different field obtained from the fits.
Inset: The Luttinger exponent $\eta$ (=~$\alpha-1$) as a function of field.
}
\end{figure}

With fields ranging from 2.2~T to 3~T, $1/T_1$ exhibits a dip feature at temperature between
3~K and 10~K (Fig.~\ref{invt16}(a)), which should indicate a dimensional crossover behavior.
 At high temperatures, the phase is characterized  by the 1D, Haldane gapped behavior, while at low temperatures, the
phase is affected by the 3D coupling which enhances spin fluctuations
and leads to AFM ordering.
Therefore a 1D QCP is hidden within the 3D ordering regime.
In the following analysis, we will determine the 1D QCP
 using the high-temperature $1/T_1$ data, where a TLL phase is established.

 In a Haldane chain system without in-plane anisotropy, the suppression of the Haldane gap by an external field should lead to a TLL state characterized by gapless excitations~\cite{2007_PRB_Takahumi,2008_PRL_Giamarchi}.
In the TLL state, the correlation function exhibit power-law decays with both a transverse and a longitudinal mode, which is written as
\begin{equation}
\begin{aligned}
&<S_0^xS_r^x>\sim(-1)^{r}r^{-\eta_x},\\
&<S_0^zS_r^z>-M^2\sim \cos(2K_F r)r^{-\eta_z}.
\end{aligned}
\end{equation}
 In the TLL state, there is a relation $\eta_z\eta_x=1$,
 where $\eta$ are the Luttinger exponent~\cite{2004_PRL_Maeshima}.
Consequently, the spin-lattice relaxation rates are expected to follow a power-law temperature dependence, given by
$1/T_1 \approx cT^\alpha$, where $\alpha=\eta-1$  and $\alpha=1/\eta-1$  for transverse and longitudinal
AFM fluctuations, respectively~\cite{2015_PRB_klanjsek}.

This is indeed can be observed through the behavior of $1/T_1$ at temperatures far above $T_{\rm N}$, where the interchain coupling $J'$ is not effective.
In the temperature regime just below 30~K, 1$/T_1$ exhibits a power-law temperature dependence, 1$/T_1$~$\sim$~$T^\alpha$,
as evidenced by the straight fit lines for fields ranging from 3.5~T and above. The power-law exponents $\alpha$ are plotted as a function of field, starting from zero at 3.5~T
and monotonically decreasing to $-$0.9 at 25~T, clearly demonstrating the presence of gapless excitations.

For the Haldane model with $D>0$, when a field is applied along the hard-axis ($b$ direction in NiCO), transverse correlations
should  dominate the low-energy spin fluctuations.
We then adopted $\eta=\alpha+1$ and calculated $\eta$ as a function of field, as shown in the inset of Fig.~\ref{invt16}(b).
The value of $\eta$ starts at 1 at 3.5~T, decreases monotonically as  field increases, and approaches zero at 25~T.

The emergence of  power-law scaling in $1/T_1$ supports the existence of a hidden 1D QCP at 3.5~T, marking a transition from a gapped Haldane phase to a gapless TLL phase.
The presence of gapless TLL behavior suggests that the in-plane anisotropy $E$ should be negligible.
Consequently, the high-field magnetic ordering could be effectively described by a magnetic
Bose-Einstein condensation (BEC) model.

\section{\label{S_phase}Phase Diagram and Discussions}

\begin{figure}[t]
\includegraphics[width=8.5cm]{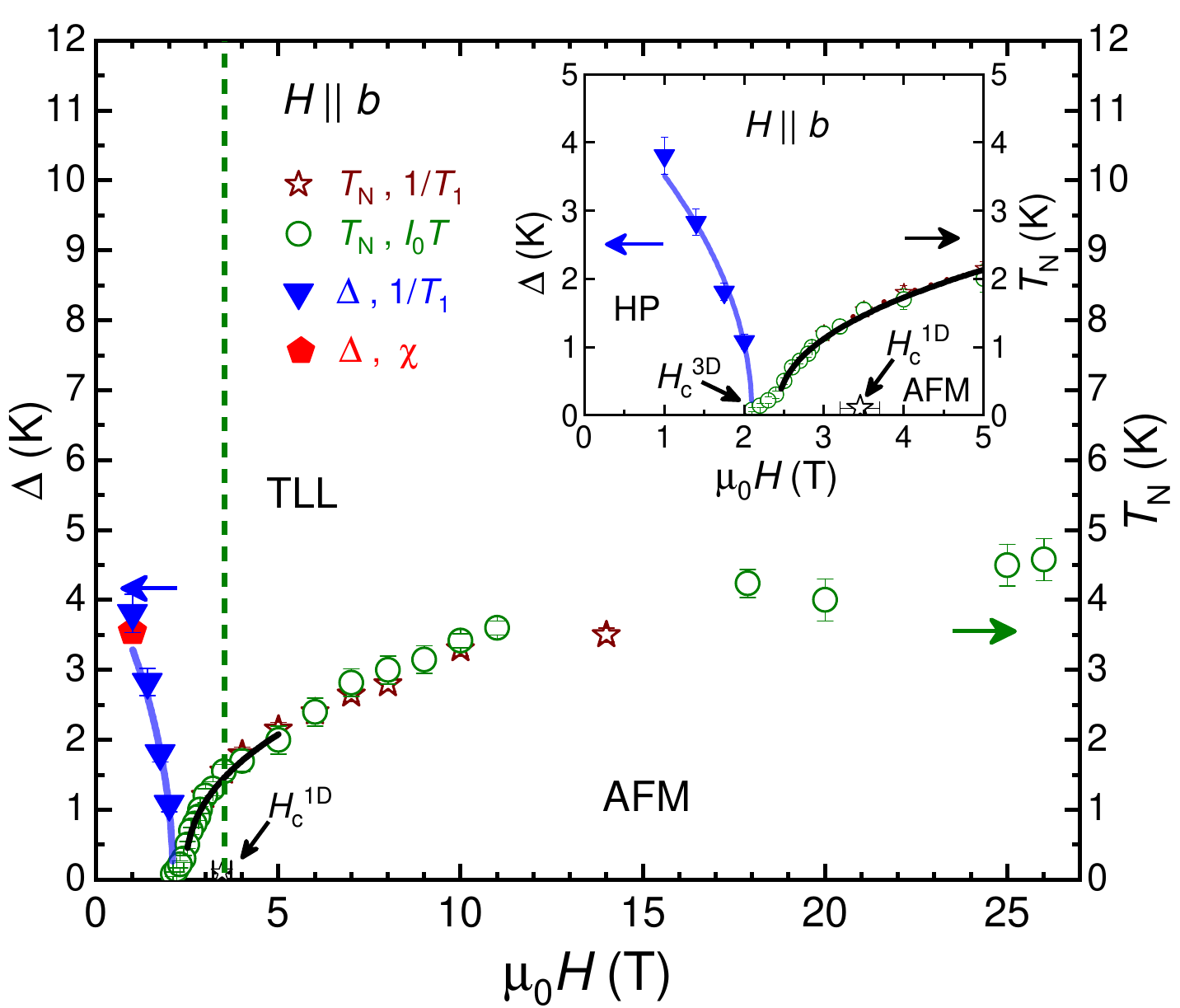}
\caption{\label{phase}
\textbf{Phase Diagram.}
The Haldane gap $\Delta$ and the phase boundaries between the Haldane phase and the ordered AFM phase determined by various probes.
 $\Delta$ intercepts the $x$-axis at $H^{\rm 3D}_{\rm c}$, and the vertical line, representing the
left boundary of the TLL phase, intercepts  the $x$ axis at the $H^{\rm 1D}_{\rm c}$.
The solid blue and black lines represent fits of $\Delta$ and $T_{\rm N}$ to power-law functions (see text).
Inset: Enlarged view of $\Delta$ and $T_{\rm N}$ at low fields.
}
\end{figure}

Our results are summarized in the $H$-$T$ phase diagram, as illustrated in Fig.~\ref{phase}.
$T_{\rm N}$ grows monotonically with field,  but only reaches 4.5~K at 26~T, which suggests a large intrachain coupling and a small interchain coupling.
Even with such a small interchain coupling, our data resolves two QCPs nearby,
that is, a 3D QCP $H^{\rm 3D}_{\rm c}$ at about 2.1~T and a hidden 1D QCP $H^{\rm 1D}_{\rm c}$ at about
3.5~T, determined by the closure of the Haldane gap at low temperatures and the onset of the TLL behavior at high temperatures, respectively.

With field far less than $H^{\rm 3D}_{\rm c}$, the interlayer coupling is negligible in gapped Haldane phase.
By extrapolation, $\Delta_{xy}=$$\Delta(H=0)$~$\approx$~4.9~K~$\approx$~$0.14J$ is obtained.
Then with Eq.~\ref{equa2}, $D~{\approx}~0.47J$ is again estimated, consistent with that obtained
by the susceptibility data.
With above data, the determined parameters in the current study are listed in Table~\ref{tab:1}.
From which, we can conclude that the NiCO exhibits an easy-plane anisotropy, with single-ion anisotropy parameter
$D>0$ and $E~{\approx}~0$.

\begin{table}[h]
 \centering
\caption{The parameters obtained by different measurements in this work.
}
\label{tab:1}
\begin{tabular}{cccccc}
 \hline\hline\noalign{\smallskip}
 ~ & ~$J$~ ~ & ~$\Delta_{xy}$~ & ~$\Delta_{xy}$~ & ~$D$~\\
       \noalign{\smallskip}\hline\noalign{\smallskip}
        &~35~K ~~ & ~3.55~K ~~& 4.9~K &~~16.58~K  \\
       \noalign{\smallskip}\hline\noalign{\smallskip}
       Methods[field] & $\chi$[1T]& $\chi$[1T]&~$1/T_1$[0T]~&~$1/T_1$\\
      \noalign{\smallskip}\hline
   \end{tabular}
\end{table}

In principle, the $T_{\rm N}$ of the AFM order should follow a power-law scaling $T_{\rm N}$$\sim$$(H^{\rm AFM}_{\rm C}-H)^{\beta}$.
As shown by the fits (black solid lines)
in the inset of Fig.~\ref{phase}, we dertermine with $H^{\rm AFM}_{\rm c}$$\approx$~2.439~T and $\beta = 0.40682$,
which is consistent with a 3D QCP of BEC~\cite{2005_PRL_Takayama}. However, we observe a deviation
from the fitting at fields below 2.44~T, as depicted in the enlarged view of the phase diagram in
the inset of Fig.~\ref{phase}, which may be affected by disorder which helps to suppress the Haldane gap at low fields.

\section{Summary}

To summarize, we conducted susceptibility and NMR studies on the spin-1 chain compound NiC$_2$O$_4$$\cdot$2NH$_3$.
Our data reveal a gapped Haldane phase at low fields and a field-induced magnetic ordering
with field above a 3D quantum critical point $H_{\rm c}^{\rm 3D}$ of approximately 2.1~T.
Additionally, Tomonaga-Luttinger liquid(TLL) behavior is observed in the field range above a 1D quantum critical point $H_{\rm c}^{\rm 1D}$($\approx$~3.5~T), at temperatures above $T_{\rm N}$.
With analysis data of the susceptibility and NMR data, we estimated the dominant intrachain exchange coupling to be approximately $J\approx$~35~K, with an easy-plane single-ion anisotropy of $D~{\approx}$~0.47~$J$
and a negligible in-plane anisotropy $E$.

\section{Acknowledgements}
The authors thank Professors Fenghua Ding, Kent Griffith and Kenneth Poeppelmeier for communicating their initial synthesis, structural and magnetic results.
This work is supported by the National Key R$\&$D Program of China (Grant No.~2023YFA1406500, No.~2022YFA1402700, No.~2023YFA1406100 and No.~2022YFA1403402),
the National Natural Science Foundation of China (Grant No.~12134020, No.~12374156, No.~12104503, No.~12374142 and No.~12304170), and the Strategic Priority Research Program(B) of the Chinese Academy of Sciences(Grant No.XDB33010100).
 A portion of this work was carried out at the Synergetic Extreme Condition User Facility (SECUF).


%

\end{document}